\documentclass[letter]{./aa}
\usepackage{graphicx,epsfig,fancyhdr,rotating,amsmath,natbib}
\usepackage{txfonts,url}
\def\etal {{ et al. }}

\usepackage{txfonts}
\begin{document}

\title{Propagating waves in polar coronal holes as seen by  SUMER \&\ EIS }
\author{D. Banerjee\inst{1} \and L. Teriaca\inst{2} \and G. R. Gupta\inst{1,3}
\and S. Imada\inst{4} \and G. Stenborg\inst{5} \and S. K. Solanki\inst{2,6}}

\institute{Indian Institute of Astrophysics, Bangalore 560034, India
\and Max-Planck-Institut f\"{u}r Sonnensystemforschung (MPS), 37191
     Katlenburg-Lindau, Germany
\and Joint Astronomy Programme, Indian Institute of Science, Bangalore 560012, India
\and National Astronomical Observatory of Japan, 2-21-1 Osawa, Mitaka, Tokyo 181-8588, Japan
\and Interferometrics, Inc., Herndon, VA 20171, USA
\and School of Space Research, Kyung Hee University, Yongin, Gyeonggi 446-701,
Korea}

\date{}

\abstract {To study the dynamics of coronal holes and the role of waves in
the acceleration of the solar wind, spectral observations were performed over
polar coronal hole regions with the SUMER spectrometer on SoHO and the EIS
spectrometer on Hinode.}
{Using these observations, we aim to detect the presence of propagating waves in the corona and to study their
properties.}
{The observations analysed here consist of SUMER spectra of the
Ne~{\sc viii} 770 \AA\ line (T~$=$~0.6~MK) and EIS slot images in the
Fe~{\sc xii}~195~\AA\ line (T~$=$~1.3~MK).
Using the wavelet technique, we study line radiance oscillations at
different heights from the limb in the polar coronal hole regions.}
{We detect the presence of long period oscillations with periods of 10 to 30
min in polar coronal holes.
The oscillations have an amplitude of a few percent in radiance and are not
detectable in line-of-sight velocity. From the time distance maps we find
evidence for propagating velocities from 75 km~s$^{-1}$ (Ne~{\sc viii}) to
125 km~s$^{-1}$ (Fe~{\sc xii}). These velocities are subsonic and roughly in 
the same ratio as the respective sound speeds.}
{We interpret the observed propagating oscillations in terms of slow
magneto-acoustic waves.
These waves can be important for the acceleration
of the fast solar wind.}
\authorrunning{Banerjee \etal}
\titlerunning{Propagating waves in the corona}
\maketitle
\section{Introduction}
Propagating radiance oscillations were detected in polar plumes,
first by \cite{1997ApJ...491L.111O} using
UVCS/SoHO and later by
\cite{1998ApJ...501L.217D} with EIT/SoHO.
\cite{1999ApJ...514..441O, 2000ApJ...533.1071O} identified the observed
radiance oscillations as propagating slow magneto-acoustic waves.
A number of studies using the CDS/SoHO spectrometer have
reported oscillations in the polar coronal holes up to 25\arcsec above 
the limb
\citep[e.g.,][]{2000SoPh..196...63B,
2001A&A...377..691B, 2001A&A...380L..39B, 2006A&A...452.1059O,
2007A&A...463..713O}.
\citet{2005A&A...442.1087P}, using SUMER/SoHO, detected radiance
fluctuations with periods from 10 to 90 min up to 15\arcsec\ above the limb.
These studies point to the presence of compressional waves, thought to
be slow magneto-acoustic waves.
In this letter, for the first time to our knowledge, 
simultaneous use of the SUMER and EIS/Hinode spectrometer were used to study these 
propagating disturbances in the off-limb
regions of the polar coronal holes. We construct time distance maps to
study the properties of wave propagation and use wavelet analysis to
establish their periods.
Spectroscopic observations have the advantage of a narrow temperature
response (by isolating specific spectral lines) and of allowing
the study of resolved and unresolved plasma motions by measuring the
Doppler shift and width of the observed profiles. These observables
provide important constraints in establishing the nature 
of the observed oscillations.
\section{Observations}
The data analysed here were obtained on 8$^{th}$ and 15$^{th}$ April
2007 during a Hinode/SUMER joint observing campaign as part of the
Hinode Observing Programme (HOP)~45/Joint Observing program (JOP)~196. They
consist of time series taken in the south polar coronal hole by the Solar
Ultraviolet Measurements Of Emitted Radiation
\citep[SUMER,][]{1995SoPh..162..189W} spectrometer aboard the Solar and 
Heliospheric Observatory (SoHO) and by the EUV imaging spectrometer
\citep[EIS,][]{2007SoPh..243...19C} aboard Hinode
\citep{2007SoPh..243....3K}.
For SUMER, the $1\arcsec \times 120\arcsec$ slit was centred on the limb
and spectra were acquired from 19:13 to 20:47 UTC on 8$^{th}$ and from 10:44
to 14:45 UTC on 15$^{th}$, with a cadence of 18.12~s.
For EIS, the $40\arcsec$ wide slot was used to obtain
$40\arcsec\times 160\arcsec$ images in several spectral lines over the time
interval 18:42 to 20:58 UTC on 8$^{th}$ and 10:54 to 15:57 UTC on 15$^{th}$
April. The EIS data consist of a series of elementary rasters each formed by two slot
images displaced by 20$''$ in the X direction to maximise the
chances of overlapping with the other instruments. Each slot image was
exposed for about 7.5~s. As a result, for each dataset, we have two time series
with a cadence of 19.3~s that we identify hereafter with slot$_{0}$ (East) and
slot$_{1}$ (West). Before the start of the temporal series, raster images were obtained with SUMER
and EIS to allow the co-alignment of the different instruments.
\begin{figure}[!th] 
\begin{center} 
\includegraphics[width=4.0cm]{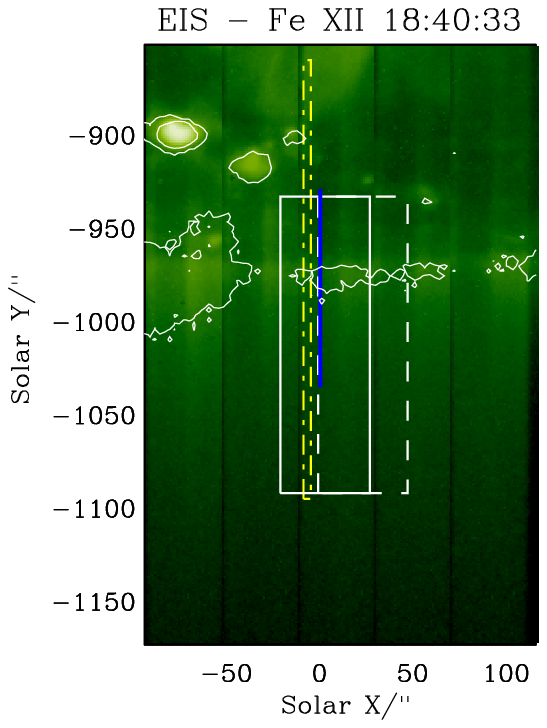}
\includegraphics[width=4.0cm]{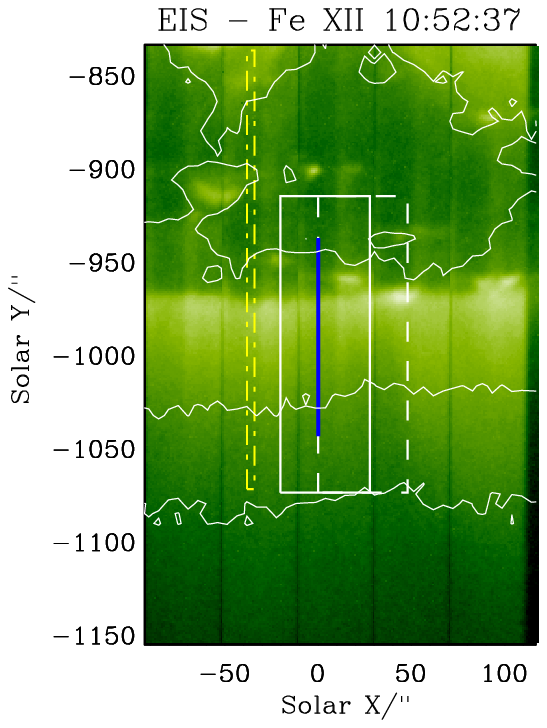}
\caption{Left (right) panel shows the location of different slits and slots on  
the Fe~{\sc xii} 195\AA\ EIS context raster on $8^{th}$ April ($15^{th}$ April). 
The context rasters are made of five adjacent slot images, covering a  
200$''\times320''$ area.
In both panels, radiance iso-contours (white) are from EIT Fe~{\sc xii} 195\AA\  
images.
The two white rectangular boxes are the location of the EIS slots
(solid slot$_{0}$, dashed slot$_{1}$).
The solid vertical blue line represents the location of the SUMER
1$''$\ wide slit whereas  
the yellow dash-dot elongated box gives the location of the 4$''$\ wide CDS  
slit.} 
\label{fig:overlay} 
\end{center} 
\end{figure} 

All the data were reduced and calibrated with the standard procedures
given in the SolarSoft
(SSW)\footnote{\url{http://sohowww.nascom.nasa.gov/solarsoft/}} library.
SUMER data were decompressed, corrected for response inhomogeneities
(flat-field) and for geometrical distortion (de-stretch).
The data series of 8$^{th}$ April was analysed only until 20:47 UTC due to
spurious flaring at the edges of the detector occurring after this time.
For EIS, after applying the standard reduction and calibration provided by the
eis$_{-}$prep procedure, data were corrected for the spacecraft jitter in the
Y-direction (the jitter in the X direction is less than 1$''$ and can be
neglected) by using housekeeping data. Finally, the movement of the
slot image on the detector due to thermal variations during the orbit was
corrected. The displacement in the dispersion (Solar X) direction was obtained
by measuring the position of the edge of the Fe~{\sc xii}~195 slot
image over time. The displacement in the Y
direction is taken equal to 1.5 times that in the X direction (Imada, 2009 in
preparation). The validity of the latter assumption was verified by checking 
the limb Y position {\it vs} time.
All images from Hinode were converted to SoHO view (L1) and co-aligned in two
steps. First, the internal offset between the long-wavelength and
short-wavelength EIS CCDs and the wavelength dependent inclination of the
spectrum were obtained (Kamio \&\ Hara, private communication) and accounted
for. Then, choosing SUMER as the reference, the EIS rasters were aligned with
the SUMER rasters. 
We estimate the alignment to be accurate within $2\arcsec$.
Fig.~\ref{fig:overlay} shows the location of the slits and slots of the
different instruments.
Here the white rectangular boxes represent EIS slot$_{0}$ and slot$_{1}$.
The vertical lines indicates the location of the SUMER (solid blue) and CDS 
(dash-dotted yellow) slits.
Due to the marginal overlap with the other instruments and the low signal in
the off-limb spectra, CDS data were not used for the analysis.
SUMER data overlap with EIS slot$_{0}$ in both cases and only those data are
used here.
\begin{figure}[!th]
\centering
{\includegraphics[width=9cm, clip=true]{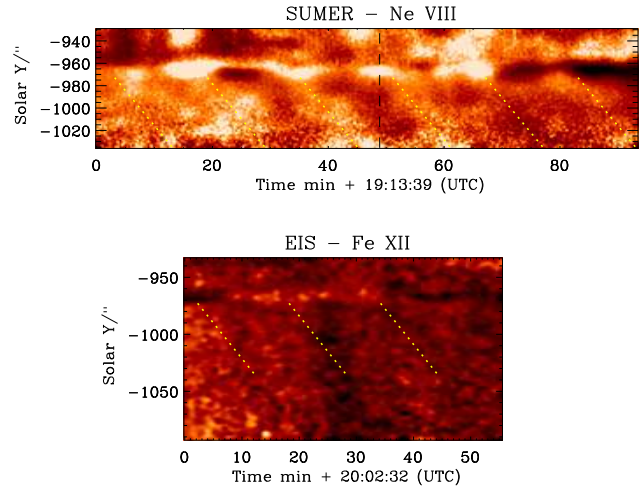}}\\
\caption{Enhanced (see text) maps of radiance variation along the slit 
(Solar Y direction) with time for Ne~{\sc viii} as recorded by SUMER 
(top panel) and Fe~{\sc xii} $195$ \AA\ as recorded by EIS (bottom panel) on 
$8^{th}$ April 2007.
The vertical black dashed line on the SUMER
enhanced radiance map depicts the starting point of the EIS time series (shown 
in the bottom panel). The slanted dotted yellow lines correspond to
disturbances propagating with a speed of 75~km~s$^{-1}$ and a period of about
15 minutes, as determined from the SUMER data.}
\label{fig:XT8aprl}
\end{figure}
\begin{figure}[!ht]
\centering
{\includegraphics[width=8.5cm, clip=true]{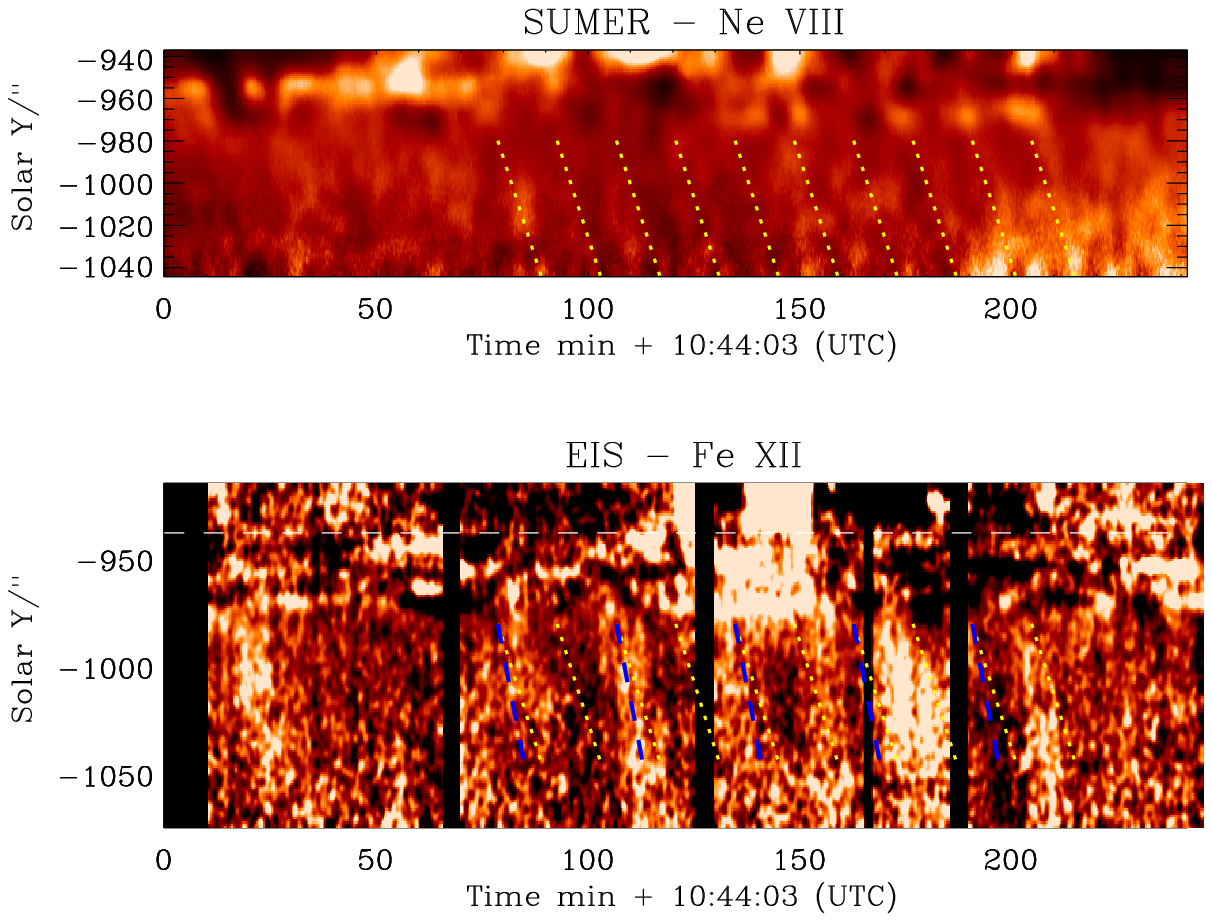}}\\
\caption{Enhanced radiance maps as in Fig.~\ref{fig:XT8aprl}, but for the
$15^{th}$ April data. Black vertical stripes correspond to data gaps in
the EIS time series. The horizontal white dashed line on the EIS
enhanced radiance map depicts the upper edge of the SUMER FOV.
The slanted dotted yellow lines correspond to
disturbances propagating with a speed of 75~km~s$^{-1}$ and a period of about 
15 minutes as determined from the SUMER data. The slanted dashed blue lines on
the EIS map correspond to disturbances propagating with a speed of
$\sim$125~km~s$^{-1}$ and a period of about 30 minutes as determined from the
EIS data.}
\label{fig:XT15aprl}
\end{figure}
%
\section{Results}
Maps of the radiance along the slit {\it vs} time (xt slices) were first built
using the SUMER Ne~{\sc viii} integrated line radiances and EIS Fe~{\sc xii}
radiances averaged over 5$''$\ in the X direction at the position overlapping
the SUMER slit. The resulting maps were then contrast enhanced and low-pass
filtered to produce the maps shown in Figs.~\ref{fig:XT8aprl}
and~\ref{fig:XT15aprl} for the $8^{th}$ and $15^{th}$ April datasets,
respectively.
EIS radiance maps were normalised by dividing by the average along the Y
direction (to remove the strong limb signature),
the result elevated to the power of
3 (to increase the contrast between high/low radiance regions),
and, finally, low-pass filtered. SUMER enhanced radiance
maps were obtained as the median over 3$\times$3 pixels of the quantity 
$exp(a+0.15\times b)^2$\ where $b$ is the original image, and a is a highly 
smoothed version of the original image obtained by convolving with an isotropic 
kernel.
The difference in the treatment of the data from
the two instruments depends on the different noise level and image scale.
In Fig.~\ref{fig:XT8aprl}, we note that the EIS time series starts at a later
time (denoted by a dashed vertical line in the top panel of the figure).
On xt slices, the presence of alternate bright and dark regions indicates the
presence of oscillations. Moreover, diagonal radiance enhancements (marked
by dotted yellow lines and dashed blue lines in Figs.~\ref{fig:XT8aprl}
and~\ref{fig:XT15aprl}) are a signature of propagating disturbances.
Thus, from such maps, it is possible to estimate periods and projected
propagation speeds.
\begin{figure*}[!t]
\centering
{\includegraphics[angle=90, width=9cm]{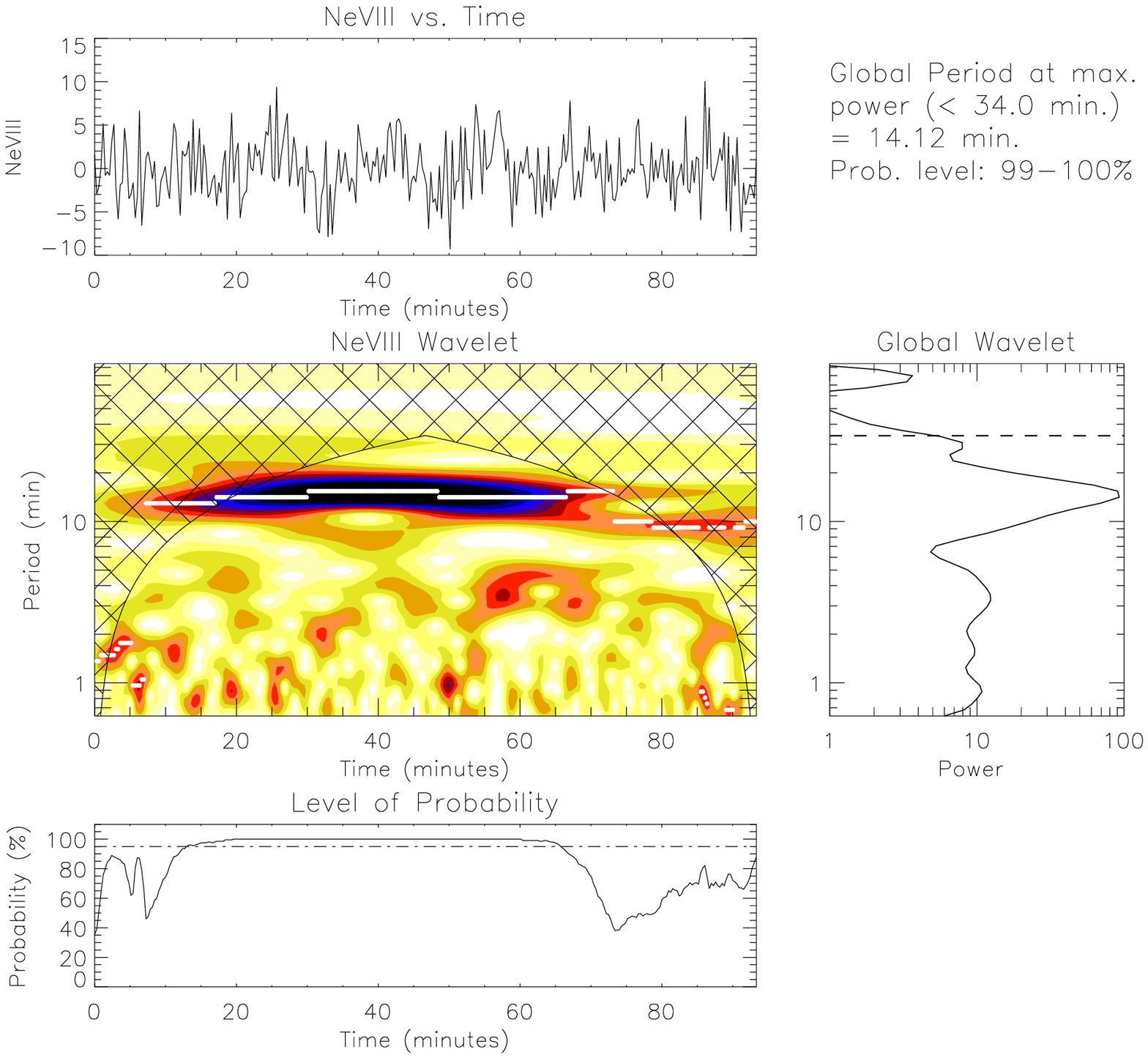}} 
{\includegraphics[angle=90, width=9cm]{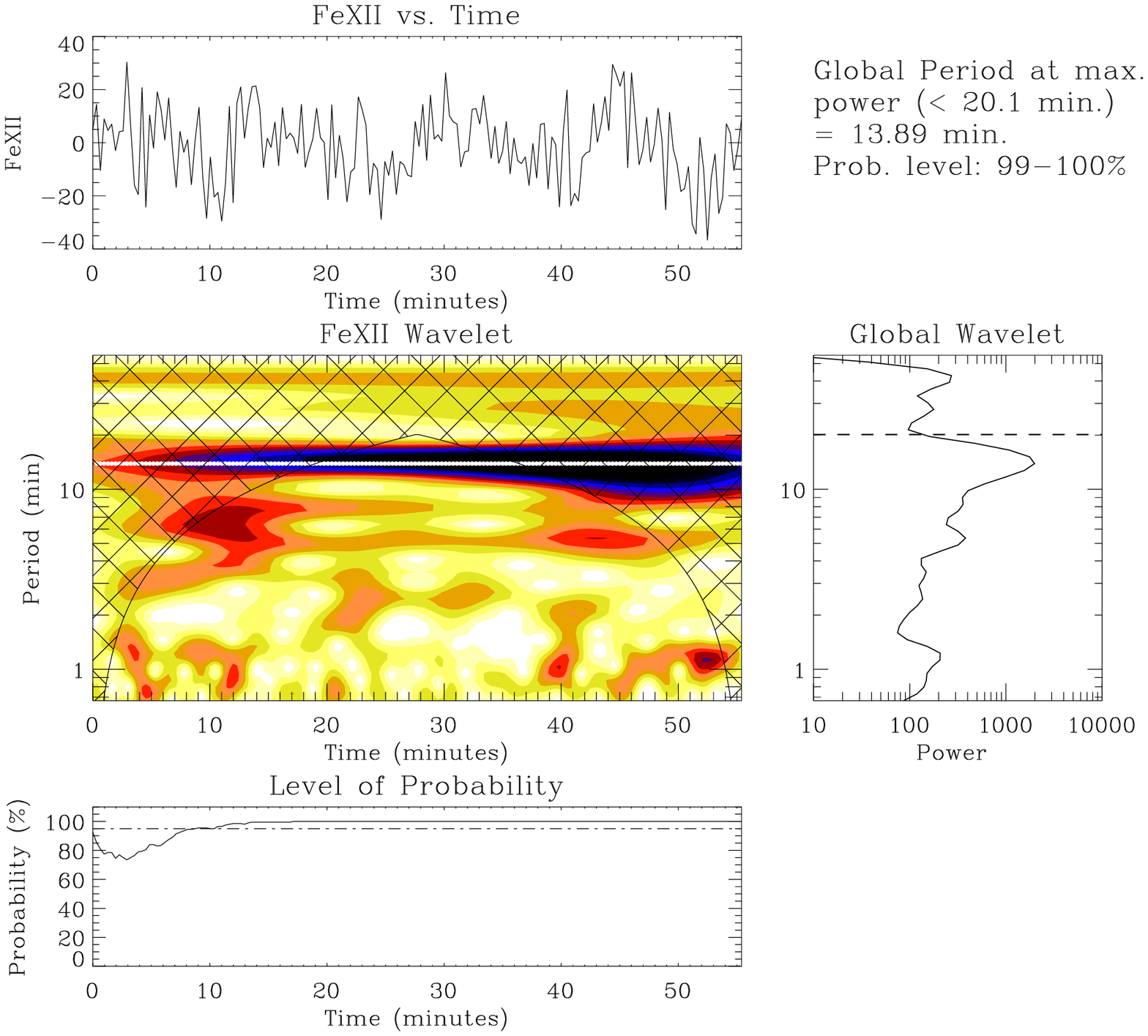}}
 \caption{Wavelet analysis for the $8^{th}$ April data at
 Solar Y$\sim-995\arcsec$ for both Ne~{\sc viii} 770~\AA\ (left) and
 Fe~{\sc xii} 195~\AA\ (right). Both light curves are obtained by averaging
 over 9$''$\ along the slit.
 In each set we show the relative (background trend-removed)
 radiance (top panels), the colour inverted wavelet power
 spectrum (central panels), the variation of the probability
 estimate associated with the maximum power in the wavelet power
 spectrum (marked with white lines) (bottom panels), and the
 global (averaged over time) wavelet power spectrum (right middle 
 panels). Above the global
 wavelet, the period (measured from the maximum power of the global wavelet),
 and the probability estimate, are given.}
 \label{fig:wavelet8april} 
\end{figure*}
\begin{figure*}[!t]
\centering
{\includegraphics[angle=90,width=9cm]{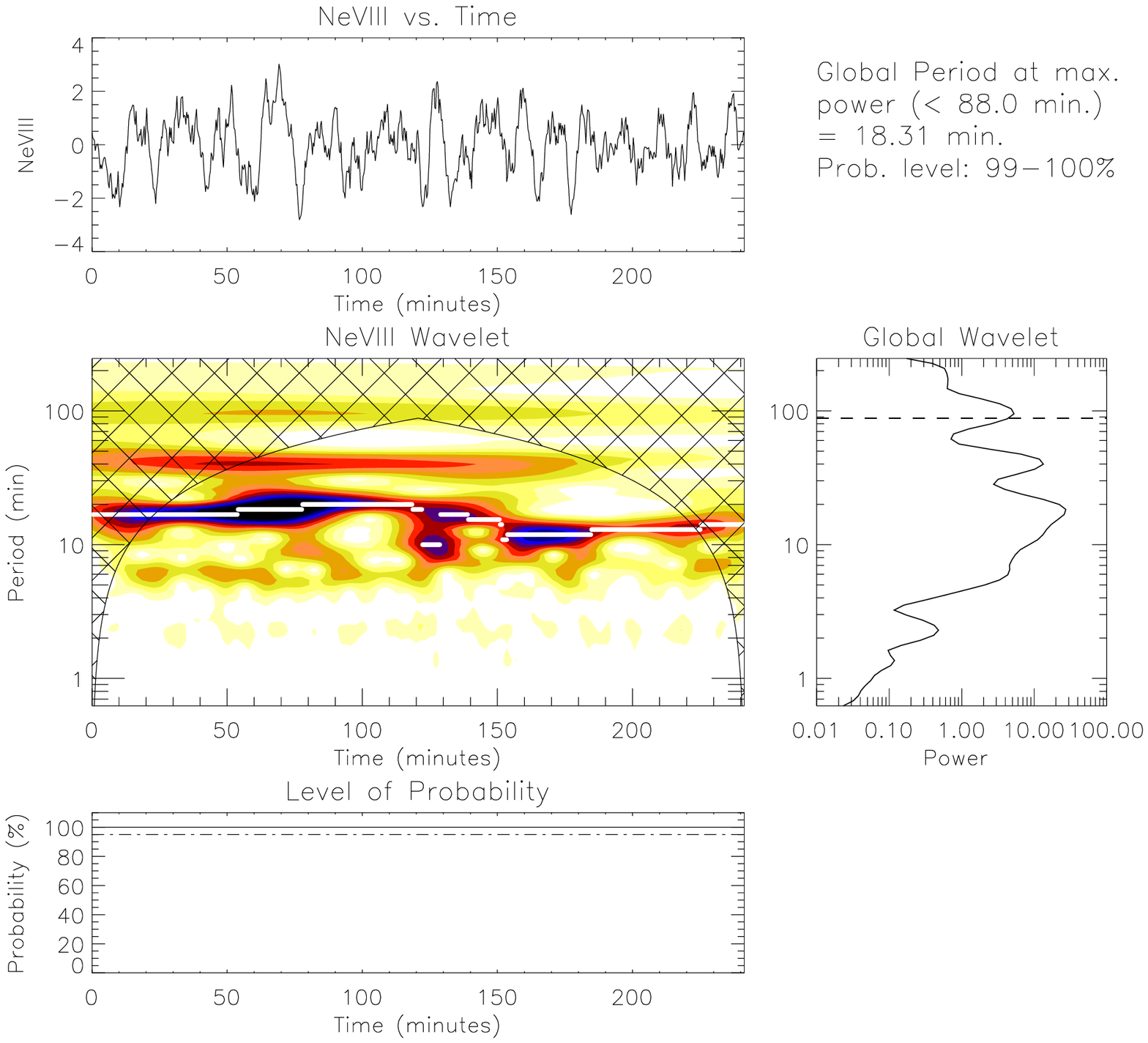}}
\includegraphics[angle=90, width=9cm]{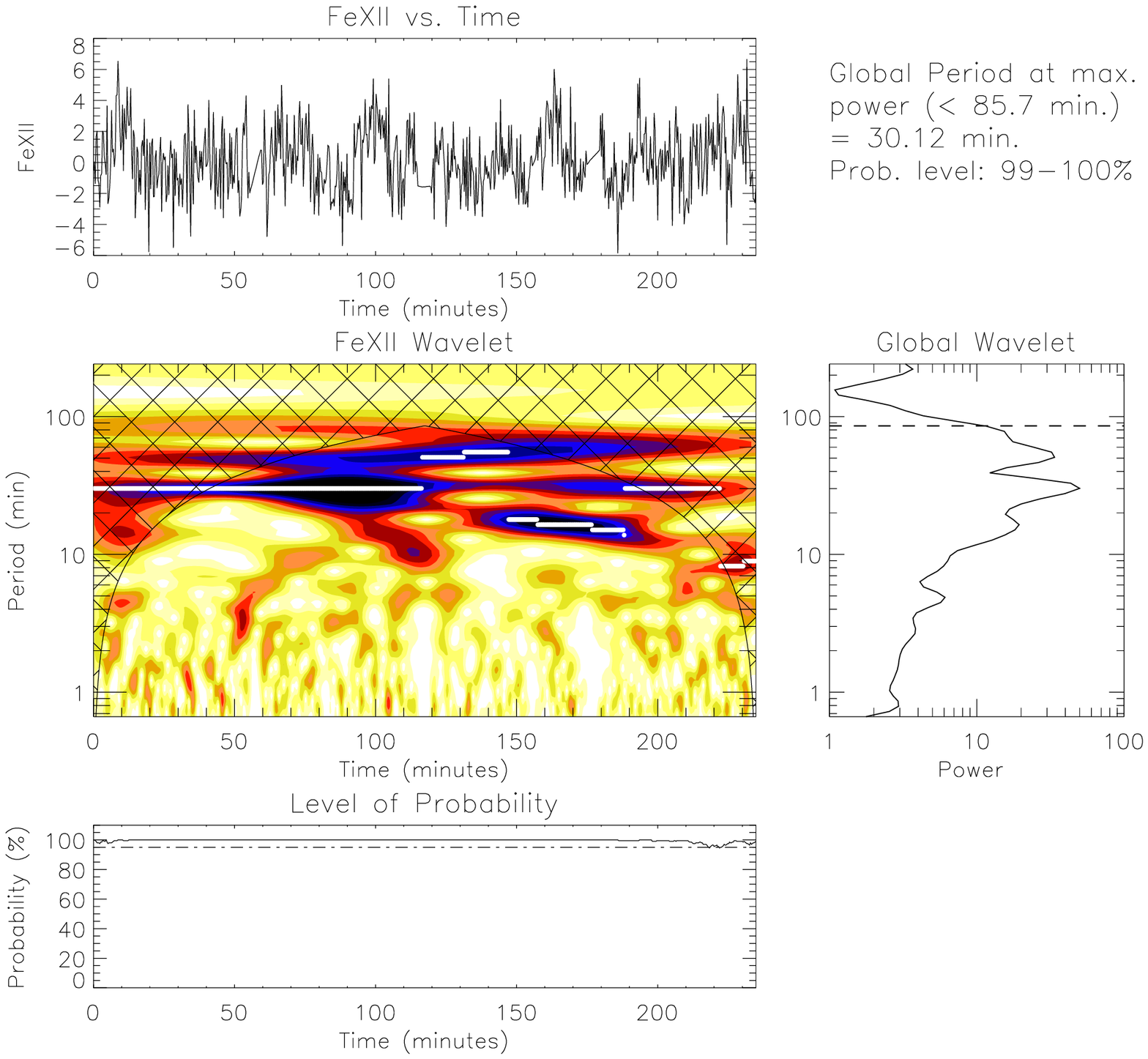}
 \caption{Wavelet analysis for the $15^{th}$ April data at
 Solar Y$\sim-995\arcsec$ for both Ne~{\sc viii} 770~\AA\ (left) and
 Fe~{\sc xii} 195~\AA\ (right). Both light curves are obtained by averaging
 over 9$''$\ along the slit. Panels as in Fig.~\ref{fig:wavelet8april}}
 \label{fig:wavelet15april}
\end{figure*}
\onlfig{6}{
\begin{figure*}[h]
\centering
{\includegraphics[angle=90,width=9cm]{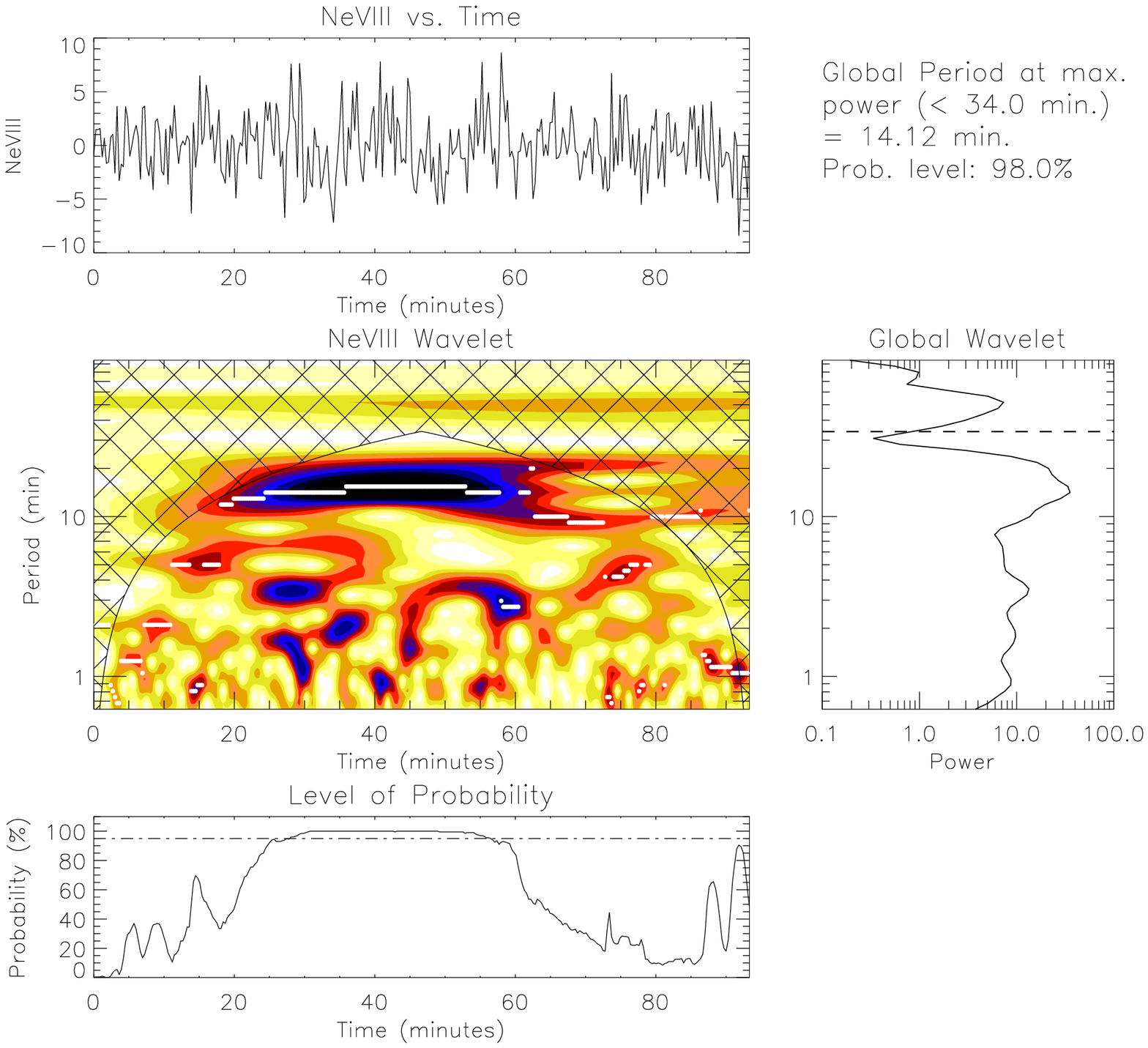}}
\includegraphics[angle=90, width=9cm]{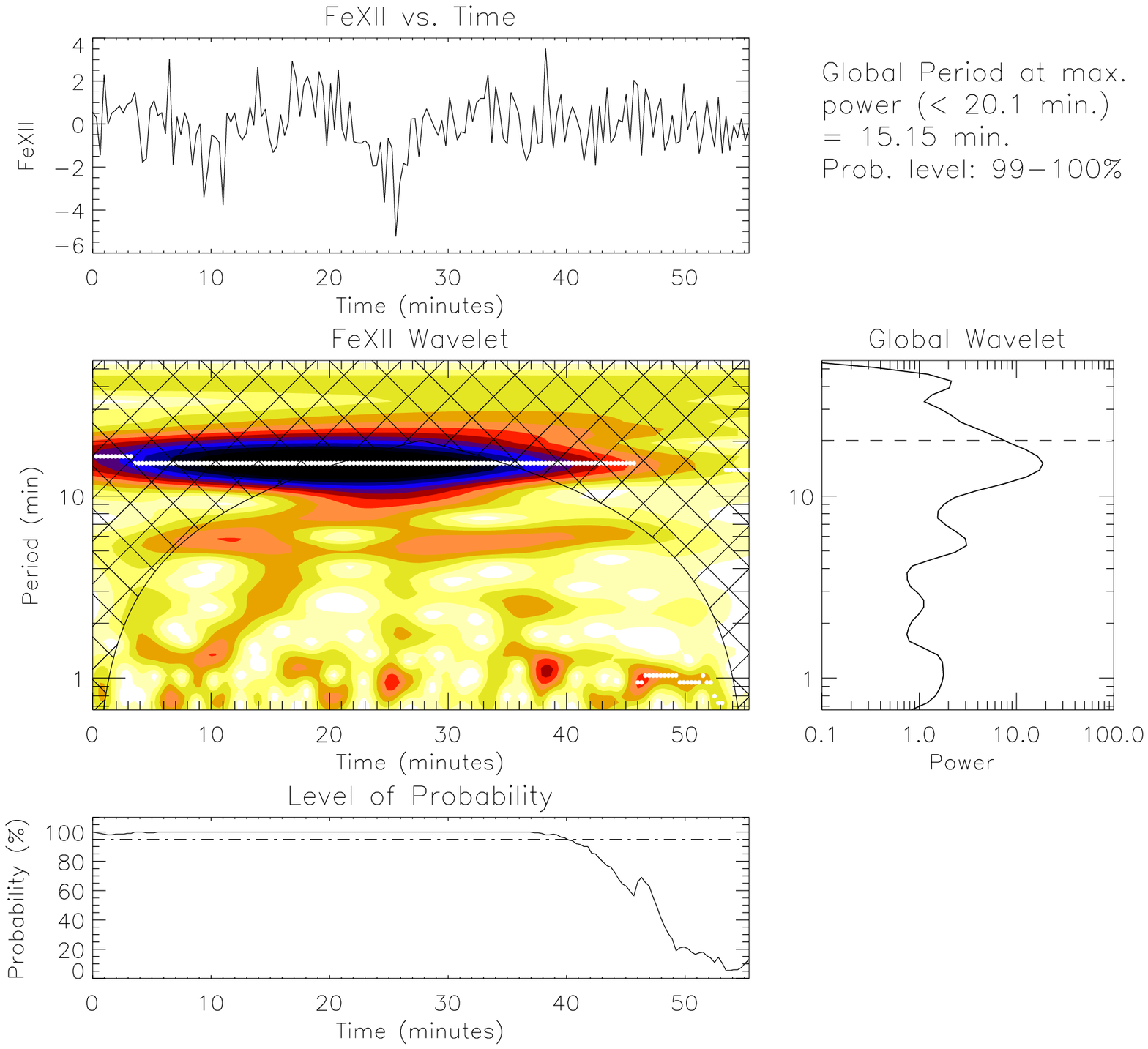}
 \caption{Wavelet analysis for the $8^{th}$ April data at
 Solar Y$\sim-1003\arcsec$ for both Ne~{\sc viii} 770~\AA\ (left) and
 Fe~{\sc xii} 195~\AA\ (right). Both light curves are obtained by averaging
 over 9$''$\ along the slit. Panels as in Fig.~\ref{fig:wavelet8april}}
 \label{fig:wavelet8april_1003}
\end{figure*}
}
\onlfig{7}{
\begin{figure*}[h] 
\centering 
{\includegraphics[angle=90,width=9cm]{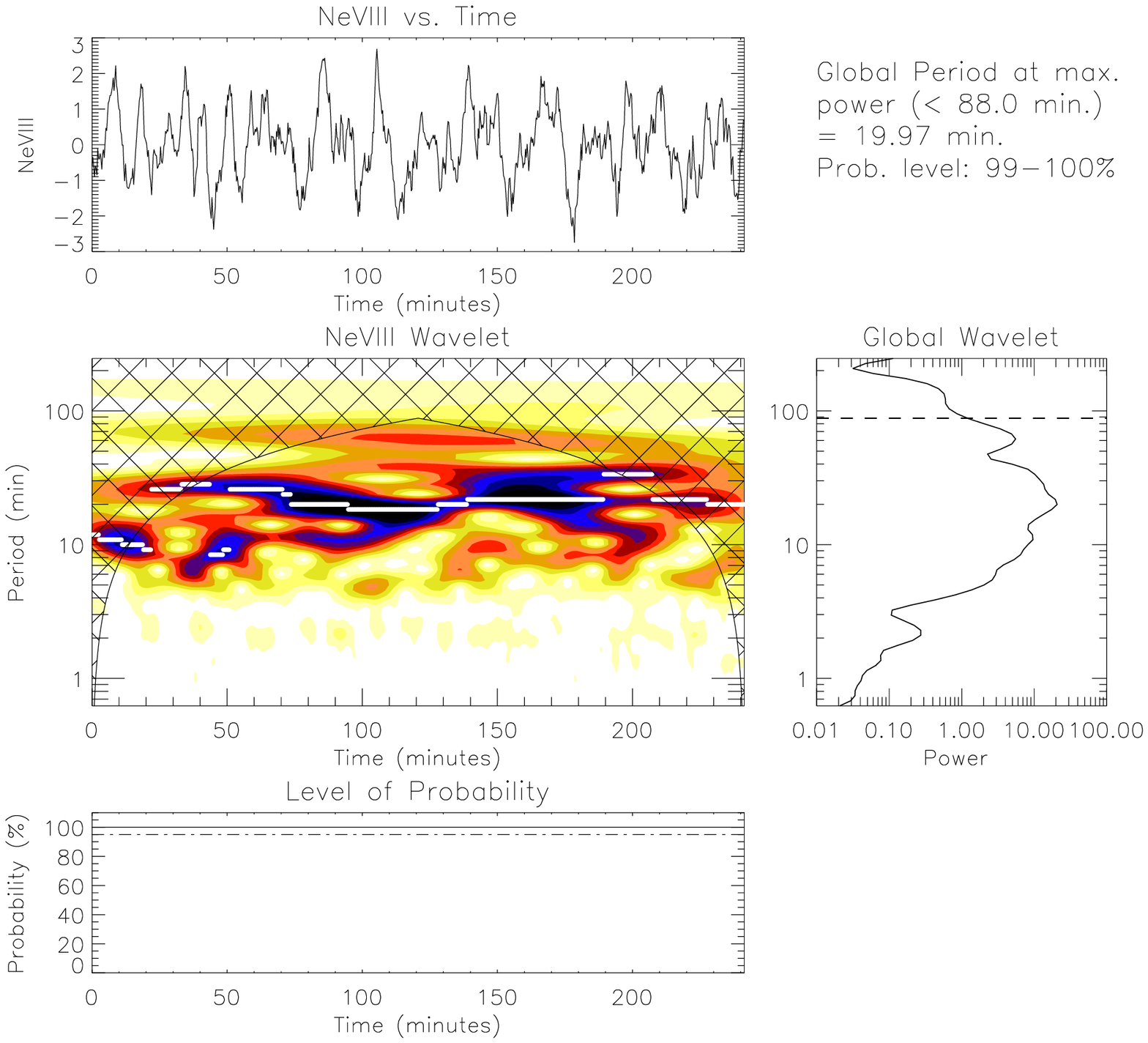}}
\includegraphics[angle=90, width=9cm]{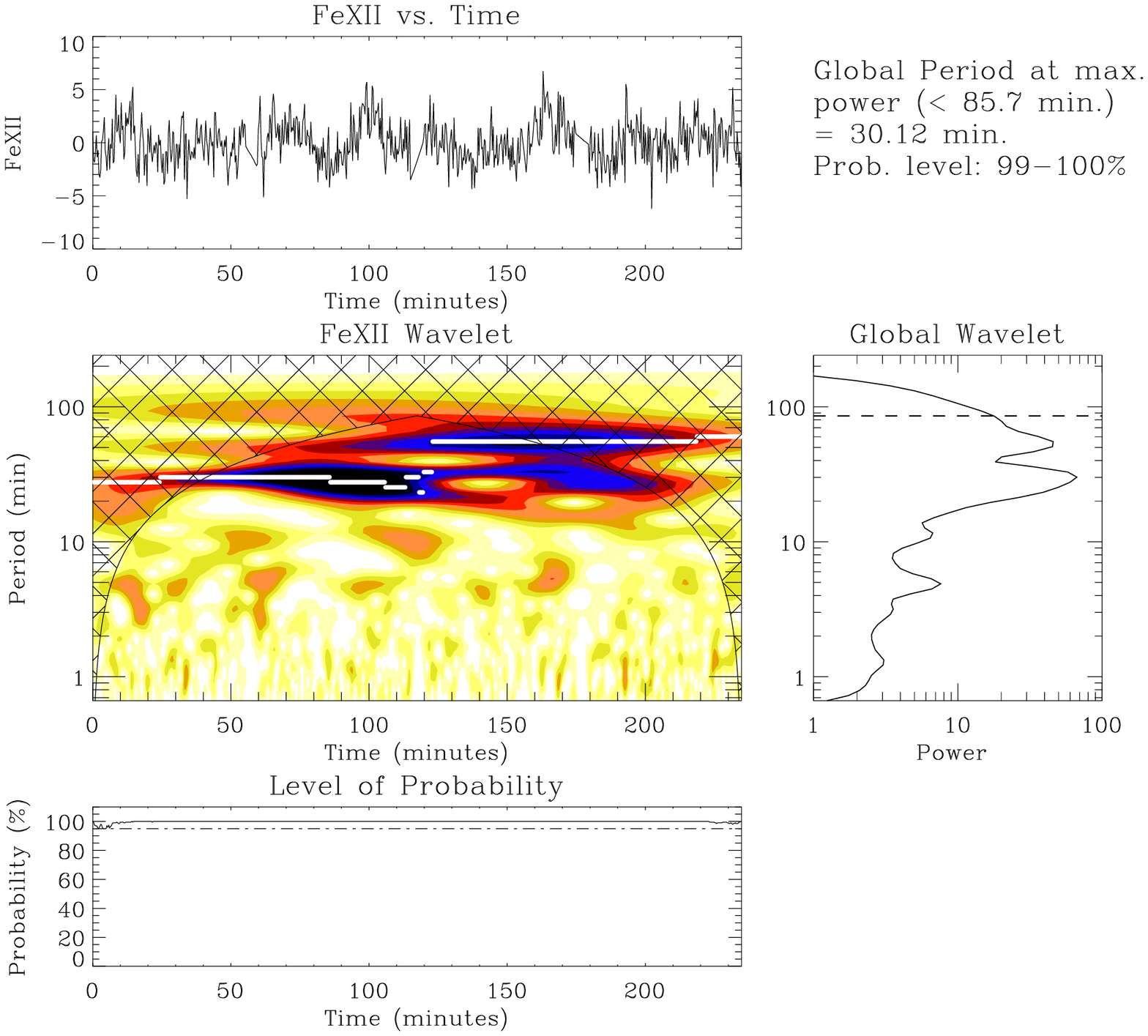}
 \caption{Wavelet analysis for the $15^{th}$ April data at
 Solar Y$\sim-1003\arcsec$ for both Ne~{\sc viii} 770~\AA\ (left) and
 Fe~{\sc xii} 195~\AA\ (right). Both light curves are obtained by averaging
 over 9$''$\ along the slit. Panels as in Fig.~\ref{fig:wavelet8april}}
 \label{fig:wavelet15april_1003}
\end{figure*}
}
From the SUMER data on both dates we measure a propagation speed of
$\sim$75~km~s$^{-1}$ and a periodicity of about 15 minutes. The propagation
fronts are much more visible on the 8$^{th}$\ April data (see
top panel of Fig.~\ref{fig:XT8aprl}) but are also visible on the
15$^{th}$\ April data. In the case of the EIS data, no definitive conclusions can be drawn from
 the 8$^{th}$\ April data (see bottom panel of Fig.~\ref{fig:XT8aprl}), 
while propagating fronts are clearly visible in the
15$^{th}$\ April data (see bottom panel of Fig.~\ref{fig:XT15aprl}).
However, both the propagation speeds and the periodicity are higher than for
the SUMER data, with values of $\sim$125~km~s$^{-1}$ and 30 minutes,
respectively (see the dashed blue lines in the bottom panel of
Fig.~\ref{fig:XT15aprl}). The observed propagation speeds are actually
a lower limit as the field lines may form an angle with the
plane-of-sky. However,
these angles are very likely small and the speeds are most probably subsonic.
For the propagation speeds, we estimate the uncertainty to be around
10\%. The ratio of the
Fe~{\sc xii} to  Ne~{\sc viii} propagation speeds, about 1.7, is close
to the ratio of the sound speeds of about 1.5 (at the two formation
temperatures of 0.6 and 1.3~MK). The observed radiance perturbations have an
amplitude of a few percent.
No evidence of propagating disturbances
can be found in the LOS velocity xt maps (not shown).

To better study the properties of the propagating disturbances seen in
the enhanced-radiance xt slices in Figs.~\ref{fig:XT8aprl}
and~\ref{fig:XT15aprl}, we make use of wavelet analysis and focus
on individual locations (heights) in the off-limb corona. For this purpose we
extracted radiance {\it vs} time curves at a given height by averaging over
9\arcsec in the Y direction for both SUMER and EIS original (without filtering 
or contrast enhancement) radiance maps.
In Fig.~\ref{fig:wavelet8april} ($8^{th}$ April) and
Fig.~\ref{fig:wavelet15april} ($15^{th}$ April),
we show representative examples of the type of oscillation measured at
Y$=~-995\arcsec$\ ($\sim 28\arcsec$\ above the limb).
The top panels of Figs.~\ref{fig:wavelet8april} and~\ref{fig:wavelet15april}
show the variation of the radiance (hereafter we will use the term
radiance for trend-subtracted line radiance). Details on the wavelet
analysis, which provides information on
the temporal variation of a signal, are described in
\citet{1998BAMS...79...61T}. For the convolution with the time series in the
wavelet transform, we chose the Morlet function, as defined in
\citet{1998BAMS...79...61T}.
The light curves shown in the upper panels had
their background trend removed by subtracting from the original time series a
$60-$point ($\approx18.1$ min) running
average for $8^{th}$ April data and a $100-$point ($\approx30.2$ min) for SUMER
and $150-$point ($\approx45.3$ min) for the EIS  running average for $15^{th}$
April data. In the wavelet spectrum (middle-left
panels), the cross-hatched regions are locations where estimates of
oscillation periods become unreliable. This is the so-called cone-of-influence
\citep[COI, see][]{1998BAMS...79...61T}.
As a result of the COI, the maximum measurable period is shown by a dotted
line in the global spectrum plots (middle-right panels).
Above the global  wavelet spectrum of Figs.~\ref{fig:wavelet8april} 
and~\ref{fig:wavelet15april} we show the prevalent period, measured at the location of the
maximum of the global wavelet spectrum, together with an estimate of the
probability that this oscillation is not due to noise. The probability estimate
was calculated using the randomisation method with 200 permutations as
outlined in detail in \cite{2001A&A...368.1095O}. 

Below the wavelet power spectrum, in the lower panels, we
show the variation of the probability estimate, calculated using the
randomisation technique, associated with the maximum power at each
time in the wavelet power spectrum. The location of the maximum power
is indicated by the over-plotted white lines in the middle-left panels.
Figure~\ref{fig:wavelet8april} shows that the oscillations detected in the 
$8^{th}$ April data have periodicities with maximum power at $\sim$14.1~min in 
Ne~{\sc viii} (SUMER) and $\sim$13.9~min in Fe~{\sc xii}
(EIS). Note that the wavelet technique reveals highly significant
power in Fe~{\sc xii} (also at other heights), although no clear oscillation
could be discerned by eye in the EIS enhanced radiance map (see bottom panel
of Fig.~\ref{fig:XT8aprl}).
On the other hand, for $15^{th}$ April data, which have
a longer time series, Ne~{\sc viii} shows its maximum power around 18 min
and Fe~{\sc xii} shows maximum power around 30 min, consistent with the
results from the analysis of the xt slices. Also note that there are multiple
peaks seen in the global wavelet spectra (right panels), which implies that the
signal could be composed of multiple modes of oscillation.
To study the oscillation behaviour at different heights we also look at
the wavelet results at
Y$=-1003\arcsec$ and plot the results in Figs.~\ref{fig:wavelet8april_1003} 
and~\ref{fig:wavelet15april_1003}, respectively (available online only).
For the $8^{th}$ April data at Y$=-1003\arcsec$, SUMER reveals a
strong peak at 14.2 min, while EIS shows a peak at 15.15 min.
For the $15^{th}$ April data at $Y=-1003\arcsec$ SUMER
shows a strong peak at 20 min, whereas EIS shows a peak at 30 min.
Finally, we also attempted to search for oscillations in the LOS
velocity (by fitting the profiles obtained using the same binning as for the
radiance wavelet analysis).
No evidence of oscillations with an amplitude larger than 2~km~s$^{-1}$
can be found in the LOS velocity data.
This is the accuracy achievable with SUMER when observing the Ne~{\sc viii} 
line in the first order of diffraction (at the Signal to Noise values of relevance here).
\section{Conclusions}
The first observational detection of longitudinal waves came from analysing
polarised brightness (density) fluctuations in white light data.
Fluctuations with periods of about 9 min were detected in coronal holes at
a height of about 1.9$R_{\sun}$\ by \cite{1997ApJ...491L.111O} using the white
light channel of UVCS/SoHO. In a follow-up study,~\cite{2000ApJ...529..592O} 
determined the
fluctuation periods to be in the range of 7 to 10 min. The propagation
speeds of the fluctuations indicated values in the range of 160 to 260
km~s$^{-1}$ at 2~$R_{\sun}$, which is slightly slower than the acoustic speed 
at those heights. \cite{1998ApJ...501L.217D}, using EIT 171~\AA\, reported
detection of outwardly propagating radiance perturbations at distances
of 1.01 to 1.2~$R_{\sun}$, gathered in quasi-periodic groups of 3 to 10 periods,
with periods of about 10 to 15 min. The projected speeds are about 75 to
150 km~s$^{-1}$ and the relative amplitude (in density) was about 2 to 4 \%.
Usually these waves are observed propagating along the assumed coronal magnetic
structures and, thus, along the magnetic field. Their speeds are usually much
slower than the expected coronal Alfv\'{e}n speed, which leads to their
interpretation as longitudinally propagating slow magneto-acoustic waves.
Slow magneto-acoustic waves follow magnetic field lines and propagate at the
local sound speed. We detect the presence of long period oscillations with
periods of 10 to 30 min in polar coronal holes  within the range of 1 to
1.2~$R_{\sun}$, with a clear signature of propagation with velocities from
75 to 125 km~s$^{-1}$, depending on the temperature of line formation.
The measured propagation speeds are subsonic, indicating
that they are slow magneto-acoustic in nature, which is consistent with earlier
reports.
Note that this detection has been confirmed through the analysis of data from
two separate spectrometers on-board two different satellites.
Thus we feel that this simultaneous detection makes the result very robust.
 We also find that the propagation speed in Fe~{\sc xii}
($\sim$125~km~s$^{-1}$) is higher than that in Ne~{\sc viii}
($\sim$75~km~s$^{-1}$), as shown in the bottom panel of Fig.~\ref{fig:XT15aprl}.
This may be a temperature effect, as the ratio of the Fe~{\sc xii} to
Ne~{\sc viii} propagation speeds, about 1.7, is close to the ratio of the sound
speeds. Different propagating speeds observed in different
lines may also be interpreted as an indication of the presence of structures
with different temperatures along the line of sight (e.g., weak plumes).
The observed region can be either a bundle of magnetic
threads of different temperatures, or have a transverse temperature profile.
However, with the available data this statement remains a conjecture only.
Finally, we note that the observed waves have no detectable signature in the 
LOS velocity, enforcing the idea of compressive longitudinal magneto-acoustic
waves.

\begin{acknowledgements}
This work was supported by the Indo-German DST-DAAD joint project D/07/03045.
The SUMER project is financially supported by DLR, CNES, NASA, and the ESA
PRODEX programme (Swiss contribution).
Hinode is a Japanese mission developed and launched by ISAS/JAXA, with NAOJ
as domestic partner and NASA and STFC (UK) as international partners.
It is operated by these agencies in co-operation with ESA and NSC (Norway).
This work was partially supported by the WCU grant No. R31-10016 from the
Korean Ministry of Education, Science and Technology.
We thank the anonymous referee for useful comments and suggestions.
\end{acknowledgements}

\bibliographystyle{aa}
\bibliography{ban12059}
\Online
\begin{appendix}

\end{appendix}

\end{document}